\documentclass[prd,twocolumn,superscriptaddress,showpacs]{revtex4}
\usepackage{amsmath,amssymb,latexsym}
\usepackage{bm}
\usepackage{epsfig}
\usepackage{graphicx}

\newcommand{\postscript}[2]{\setlength{\epsfxsize}{#2\hsize}
   \centerline{\epsfbox{#1}}}
\newcommand{\ev}{\text{eV}}

\newcommand{\gev}{\text{GeV}}

\newcommand{\etal}{{\em et al.}}

\begin{document}

%%% Title %%%

\title{
Numerical Solutions of Inflating Higher Dimensional Global Defects
}

%%% Authors %%%

\author{Satsuki Shimono}
\affiliation{
Department of Physics,
Kyoto University, Kyoto 606-8502, Japan
}
\author{Takeshi Chiba}
\affiliation{
Department of Physics,
College of Humanities and Sciences, Nihon University,
 Tokyo 156-8550, Japan
}
\affiliation{%
Division of Theoretical Astronomy, National Astronomical Observatory of 
Japan, Tokyo 181-8588, Japan }

\date{\today}

\begin{abstract}

We find numerical solutions of Einstein equations and scalar field equation 
for a global defect in higher dimensional spacetimes ($\geq 6$). 
We examine in detail the relation among the expansion rate $H$ 
and the symmetry-breaking scale $\eta$ and the number of extra dimensions $n$ 
for these solutions. We find that even if the extra dimensions do not have 
a cigar geometry, the expansion rate $H$ grows as $\eta$ increases, which is 
opposite to what is needed for the recently proposed mechanism for solving 
the cosmological constant problem.
We also find that the expansion rate $H$ decreases as $n$ increases.
\end{abstract}


%04.50.+h   Gravity in more than four dimensions, Kaluza-Klein theory, 
%           unified field theories; alternative theories of gravity
%98.80.Cq   Particle-theory and field-theory models of the early Universe

\maketitle

\section{Introduction}

Recent observations suggest that majority of the Universe is the unknown: 
27\% is matter (and only 4\% is ordinary matter) and the rest is 
the cosmological constant (or dark energy) \cite{obs1,obs2}. 
Since the cosmological constant is interpreted as a vacuum energy \cite{zel}, 
these observations indicate that the energy density of the vacuum energy is 
$\rho_{\Lambda}\approx(10^{-3}\ev)^4$. 
On the other hand, a field theoretical estimate of zero point energy of 
the vacuum yields the Planck energy $\approx 10^{18}\gev$.
Therefore, the mismatch between observations and the theory is huge: 
${(10^{-3}\ev)^4}/{M_p^4}\simeq 10^{-120}$. 
This mismatch is called the cosmological constant problem (CCP) 
\cite{Wein2,PeebRa} and still remains to be solved.

Dvali \etal \cite{DGS} has suggested the mechanism of diluting the 
cosmological constant by using brane world models with codimension greater 
than 2. In this mechanism the observed effective cosmological constant 
related to the expansion rate becomes small 
because whose energy is consumed to bend the bulk space
even if the bare vacuum energy (brane tension) is as large as the Planck scale.
The success of the mechanism rests on the conjectured relation in higher 
dimension
\begin{equation}
H \simeq M_* \left(\frac{M_*^4}{\rho_{\Lambda 4}}\right)^{1/(n-2)},
\label{eq:DGSConj}
\end{equation}
where $M_*$ is the $(4+n)$-dimensional Planck mass and $\rho_{\Lambda 4}$ is 
the four dimensional (bare) vacuum energy (brane tension). 
If the number of extra dimensions $n$ is greater than $2$,
the expansion rate $H$ is inversely proportional to the brane tension
$\rho_{\Lambda 4}$. Then the smallness of the cosmological constant could be 
explained by the largeness of the brane tension. 

Cho and Vilenkin (extending the analytic solutions in \cite{ov}) 
have recently constructed numerical solutions 
of a global defect in seven ($n=3$) dimensional spacetime \cite{CV1}. 
They obtained numerical solutions of an inflating global defect if the 
symmetric breaking scale is greater than the higher dimensional Planck scale. 
Then the inflation rate is found to grow almost linearly as the brane 
tension is increased, which is opposite to what is needed to solve 
the cosmological constant problem (\ref{eq:DGSConj}).

The main purpose of this paper is to extend Cho and Vilenkin's model 
to the arbitrary $(4+n)$-dimensional one and find numerical solutions 
without ``cigar ansatz'' and to examine the relation among
the number of extra-dimensions, the energy density and the expansion rate. 

We shall find that the conjectured relation (\ref{eq:DGSConj}) does not hold 
even by using our new bulk solutions in other extra dimensions $n(\geq 2)$.
However we find that the brane's expansion rate is a monotonically 
decreasing function of the number of the extra dimensions $n$ and 
can vanish at the specific dimension.

In the following section, we introduce the model in 
Sec. \ref{sec:model}. 
Results of the numerical integration are shown in Sec. \ref{sec:numericalSol}.
Finally, we summarize our results in Sec. \ref{sec:conclusion}. 
Some numerical details are given in Appendix \ref{sec:numerical}. 
Cigar type solutions are reexamined in Appendix \ref{sec:cigar}.

\section{Model}

\label{sec:model}

\subsection{Space-time structure}

In our model, the brane is assumed a $4$-dimensional de-Sitter apace 
$\mathbf{dS}^4$, and the extra space is a spherically symmetric
$n$-dimensional space $\mathbf{R}\times \mathbf{S}^{n-1}$.
Where the number of extra dimensions $n$ is equal to or
greater than $2$. 
The entire manifold is wrapped product of both spaces
$\mathbf{R}\times\mathbf{S}^{n-1}\times \mathbf{dS}^4$,
whose metric is   
\begin{eqnarray}
ds^2 &=& dr^2+C(r)^2r^2d\boldsymbol{\Omega}_{n-1}^2 \nonumber \\
 &+& B(r)^2\left(-dt^2+e^{2Ht}\sum_{i=1}^{3}{dx^i}^2\right). \label{eq:metric}
\end{eqnarray}
Here the coordinate of the brane is $(t, x^1, x^2, x^3)$ and $H$ is the 
positive constant expansion rate. The extra space's coordinate
is $(r, \theta_1,\ldots ,\theta_{n-1})$ and $d\boldsymbol{\Omega}_{n-1}$ is
the metric of an $n-1$ dimensional sphere $\mathbf{S}^{n-1}$.
$C(r)r$ and $B(r)$ are the radius of the extra space and the warp factor
depending on $r$ only.
We adopt the Einstein-Hilbert action for the space-time dynamics such that
\begin{equation}
\mathcal{S}_{\mathrm{E-H}} = \frac{1}{2\kappa^2}
\int d^{4+n}x\sqrt{-g}\mathcal{R},
\label{action:gravity} 
\end{equation}
where $\kappa^2=1/M_*^{2+n}$ with $M_{*}$ being 
the $(4 + n)$-dimensional Planck mass. 

\subsection{Energy Momentum Tensor}

The global defect in the $n$-dimensional spherically symmetric space
is introduced to construct the brane,
which is described by a multiplet of
the scalar fields $\phi^i$ with the action,
\begin{equation}
\mathcal{S}_{\phi} = \int d^{4+n}x\sqrt{-g} \left[-\frac{1}{2}
\partial^A\phi^i\partial_A\phi_i - V(\phi)\right],
\label{action:phi}
\end{equation}
where capital letters $(A,\ldots)$ and small letters $(i,\ldots)$
run from $1$ to $4+n$ and from $1$ to $n$ respectively.
Because we consider spherically symmetric solutions only,
the scalar multiplet has been assumed to have
a hedgehog configuration, $\phi^i = \phi (r)\xi^i / r$.
Here $\phi(r)$ depends only on the radius coordinate $r$
and $\xi^i$ represent for the Cartesian coordinates of the extra space.
The potential of the scalar field $V(\phi)$ has minimum at 
$|\phi^i| = \phi = \eta$ such that
\begin{equation}
V(\phi)=\frac{\lambda}{4}(\phi^2-\eta^2)^2.
\end{equation}
The energy density due to the scalar field may be regarded as the brane tension.

\subsection{Basic Equations}

The Einstein equations and the equation of motion of the scalar field 
are derived from the action, Eq.(\ref{action:gravity}) and Eq.(\ref{action:phi}). 
The Einstein equations are
\begin{eqnarray}
{G^\mu}_\mu &=& - \frac{1}{4}\frac{~^{(4)}R}{B^2} + 3\frac{B''}{B} + 
3\left(\frac{B'}{B}\right)^2 \nonumber \\
&+&3(n-1)\left(\frac{B'}{Br}+\frac{B'C'}{BC}\right) \nonumber \\
&+&(n-1)\frac{C''}{C}+\frac{(n-2)(n-1)}{2}\left(\frac{C'}{C}\right)^2 
\nonumber \\
&+&n(n-1)\frac{C'}{Cr} + \frac{(n-2)(n-1)}{2}\left(\frac{1}{r^2}-
\frac{1}{C^2r^2}\right) \nonumber \\
&=&\kappa^2\left[-\frac{{\phi'}^2}{2}-\frac{(n-1)\phi^2}{2C^2r^2}
-\frac{\lambda}{4}\left(\phi^2-\eta^2\right)^2\right], \label{eq:EH1}
\end{eqnarray}
\begin{eqnarray}
{G^r}_r&=&6\left(\frac{B'}{B}\right)^2+4(n-1)\left(\frac{B'}{Br} + 
\frac{B'C'}{BC}\right) \nonumber \\
&+&\frac{(n-2)(n-1)}{2}\left(\frac{C'}{C}\right)^2 + 
(n-2)(n-1)\frac{C'}{Cr} \nonumber \\
&+&\frac{(n-2)(n-1)}{2}\left(\frac{1}{r^2}-\frac{1}{C^2r^2}\right)-
\frac{1}{2}\frac{~^{(4)}R}{B^2} \nonumber \\
&=&\kappa^2\left[\frac{{\phi'}^2}{2}-\frac{(n-1)\phi^2}{2C^2r^2}
-\frac{\lambda}{4}(\phi^2-\eta^2)^2\right], \label{eq:EH2}
\end{eqnarray}
\begin{eqnarray}
{G^{\theta_i}}_{\theta_i} &=& -\frac{1}{2}\frac{~^{(4)}R}{B^2} + 
4\frac{B''}{B} + 6\left(\frac{B'}{B}\right)^2 \nonumber \\
&+&4(n-2)\left(\frac{B'}{Br}+\frac{B'C'}{BC}\right) + 
(n-2)\frac{C''}{C} \nonumber \\
&+&\frac{(n-3)(n-2)}{2}\left(\frac{C'}{C}\right)^2+
(n-1)(n-2)\frac{C'}{Cr} \nonumber \\
&+&\frac{(n-3)(n-2)}{2}\left(\frac{1}{r^2}-\frac{1}{C^2r^2}\right) \nonumber \\
&=&\kappa^2\left[-\frac{{\phi'}^2}{2}-\frac{(n-3)\phi^2}{2C^2r^2}
-\frac{\lambda}{4}\left(\phi^2-\eta^2\right)^2\right]. \label{eq:EH3}
\end{eqnarray}
Here $~^{(4)}R = 12 H^2$ represents for the $4$-dimensional Ricci scalar 
depending on the expansion rate of the brane.
The prime denotes the differentiation with respect to $r$.
The equation of motion of the scalar field is
\begin{eqnarray}
\phi''&+&(n-1)\left(\frac{4}{(n-1)}\frac{B'}{B}
+\frac{C'}{C}+\frac{1}{r}\right)\phi' \nonumber \\
&-&(n-1)\frac{\phi}{C^2r^2}-\lambda\phi(\phi^2-\eta^2)=0. \label{eq:EOM1}
\end{eqnarray}
Eq.(\ref{eq:EH2}) imposes the constraint when solving eq. 
(\ref{eq:EH1}), (\ref{eq:EH3}) and (\ref{eq:EOM1})
as the second-order differential equations for $B,C$ and $\phi$.

\section{Numerical Solutions}

\label{sec:numericalSol}

%Explanation of Eigenvalue 
We have solved the Eq. (\ref{eq:EH1}), (\ref{eq:EH2}),
(\ref{eq:EH3}) and (\ref{eq:EOM1}) numerically with the initial conditions
$B(0) = C(0) = 1$, $B'(0) = C'(0) = 0$ and $\phi(0) = 0$.
These equations have a set of three parameters $(n, \kappa\eta, (\kappa/\lambda^{1/2})H)$.
It is found by the numerical integration that the proper relation among 
them is obtained under the condition that the point of a singularity 
becomes as far as possible. We assume that only 
particular combination of these parameters gives a nondiverging solution. 
We shall call such parameters eigenvalues and regard regular solutions 
with eigenvalues as physical solutions. 
Solutions obtained from parameters deviated from 
the eigenvalues have divergence in $B$ or $C$. 
Similar situations are considered in \cite{CV1} for the cigar ansatz which is 
detailed in Appendix \ref{sec:cigar}. 
We shall find yet another numerical solutions.
The set of eigenvalues forms the surface in the $3$-dimensional 
parameter space, whose shape also will be studied in the following.
The discussion of technical details for finding numerical solutions is given 
in Appendix \ref{sec:numerical}. 

\subsection{Asymptotic Solutions}

For $H = 0$, we can find an asymptotic solution analytically which is
 obtained by solving Eq.(\ref{eq:EH1}), (\ref{eq:EH2}), 
(\ref{eq:EH3}) and (\ref{eq:EOM1}) at large $r$. If $n \geq 3$, 
the solution is
\begin{equation}
\phi(\infty) = \eta, \label{eq:boundMinPhi}
\end{equation}
\begin{equation}
B^2(\infty) = \text{ constant}, \label{eq:boundMinB}
\end{equation}
\begin{equation}
C^2(\infty) = 1-\frac{(\kappa\eta)^2}{n-2}, \label{eq:boundMinC}
\end{equation}
where $(\kappa\eta)^2 \leq n-2$. {}From Eq. (\ref{eq:boundMinC}), 
the sphere $\mathbf{S}^{n-1}$ has a solid angle deficit such that
\begin{equation}
\Delta\Omega = \frac{2\pi^{n/2}}{\Gamma(n/2)}\cdot\frac{(\kappa\eta)^2}{n-2},
\end{equation}
where $\Gamma$ is the gamma function. As $\kappa\eta$ approaches
 $\sqrt{n-2}$, the deficit angle consumes the entire area.
If $n = 2$, $C(\infty)$ can take an arbitrary constant.

\subsection{$(\kappa\eta)^2 \leq n-2$ Case}

For $(\kappa\eta)^2 \leq n-2$,
a non-singular solution exists when the brane is not expanding, $H=0$.
Then, the solution takes an asymptotic form given in the last subsection.

We have solved the Einstein equations and the equation of motion of $\phi$ 
numerically  in the range of $[0,r_{\text{max}}]$. Here 
$r_{\text{max}}$ should be taken to be sufficiently large so that $\phi$ 
takes a constant value given in Eq. (\ref{eq:boundMinPhi}). The details of 
the method of numerical integration is given in Appendix \ref{sec:numerical}. 

As an example, Fig. \ref{fig:graph3065} shows a solution with the parameter
$(n, \kappa\eta, (\kappa/\lambda^{1/2})H) = (3, 0.65, 0)$.
\begin{figure}
\postscript{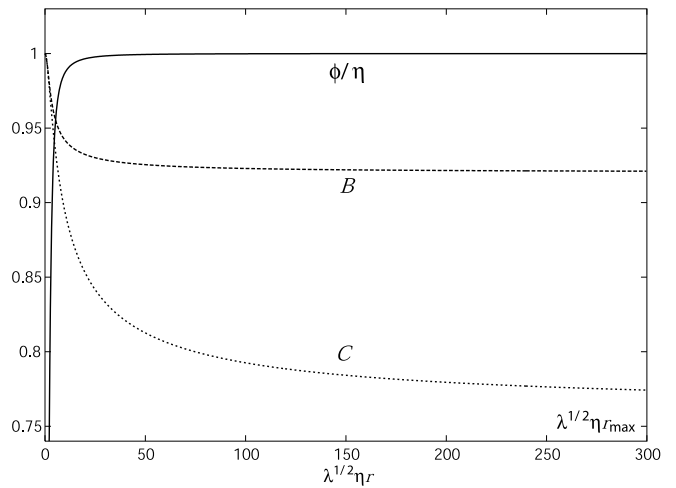}{1.00}
\caption{This graph shows a solution with the eigenvalue
$(n, \kappa\eta, (\kappa/\lambda^{1/2})H) = (3, 0.65, 0)$.} 
\label{fig:graph3065}
\end{figure}
The scalar field $\phi$ approaches $\eta$ rapidly, which makes 
the defect core with the energy density $(\kappa\eta)^4$ approximately, and
$B(r)$ and $C(r)$ approach toward constants as
Eq. (\ref{eq:boundMinPhi}), (\ref{eq:boundMinB}) and (\ref{eq:boundMinC}).

\subsection{$(\kappa\eta)^2 > n-2$ Case}

For $(\kappa\eta)^2 > n-2$, arbitrary $H$ including $H = 0$
leads to a divergence in $B$ or $C$ at
finite distance from the origin and the singularity is formed.
We call this point $r_{\text{sing}}$.
At the specific value of $H$, the distance to the singularity becomes as far 
as possible and the divergence vanishes. We call this point $r_{\text{f}}$.
We note that a fine-tuning of the parameters is required to find this 
local peak of the distance. The details of the numerical method is given 
in Appendix \ref{sec:numerical}. 

%Method of numerical integration$

\begin{figure}
\postscript{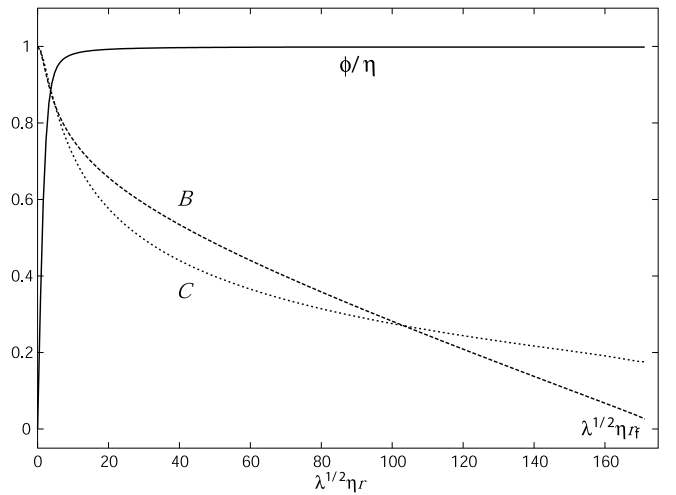}{1.00}
\caption{This graph shows the solution with the eigenvalue
$(n, \kappa\eta, (\kappa/\lambda^{1/2})H) = (3, 1.09, 0.003786056)$.
$B$ approaches toward $0$ at finite $r_{\text{f}}$.} 
\label{fig:graph3109}
\end{figure}

\begin{figure}
\postscript{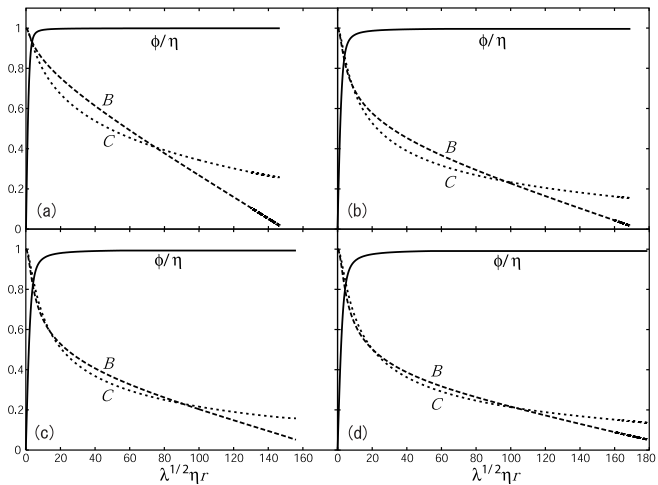}{1.00}
\caption{(a), (b), (c) and (d) are for the eigenvalues
$(n, \kappa\eta, (\kappa/\lambda^{1/2})H) = (2, 0.65, 0.003467)$, 
$(4, 1.46, 0.004330)$, $(5, 1.765, 0.004535)$ and 
$(6, 2.025, 0.004119)$ respectively.} 
\label{fig:combined}
\end{figure}

As an example, the numerical solution with the eigenvalue
$(n, \kappa\eta, (\kappa/\lambda^{1/2})H) = (3, 1.09, 0.003786056)$ is shown
in Fig. \ref{fig:graph3109}. 
It is noticed that the $B(r)$ vanishes at finite $r$ but $C(r)$ does not  
diverge. The solutions for other dimensions can be 
obtained and are shown in Fig. \ref{fig:combined}. 

\begin{figure}
\postscript{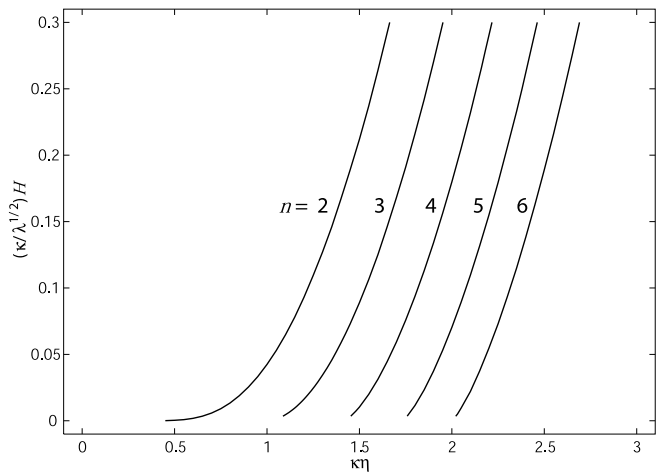}{1.00}
\caption{The relation between $\eta$ and $H$
with $n$ fixed respectively. In the region where $H$ is small, we cannot find
eigenvalue, because the point of singularity is very far from the origin.
$H = 0$ seems to be established at $\kappa\eta = \sqrt{n-2}$.} 
\label{fig:nonSingLocation23456}
\end{figure}

Fig. \ref{fig:nonSingLocation23456} shows the relations
between $\eta$ and $H$ with $n$ fixed at some values.
Each line approaches the point $(\kappa\eta, H) = (\sqrt{n-2}, 0)$. 
We find that $H$ grows as $\eta$ is increased. 
This tendency is similar to the Friedmann equation but opposite to 
the conjectured relation  Eq.(\ref{eq:DGSConj}). 

\begin{figure}
\postscript{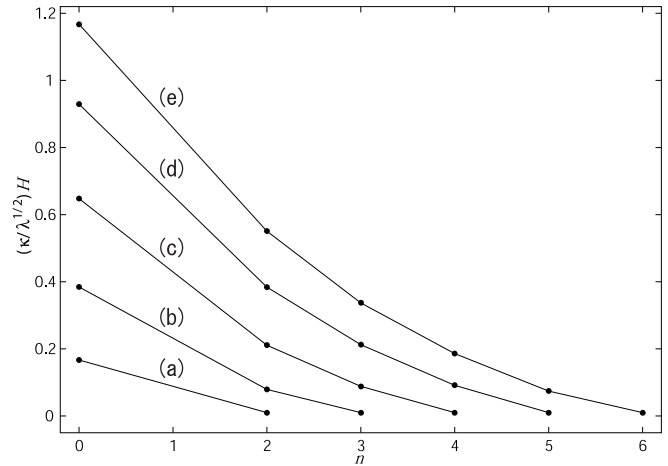}{1.00}
\caption{The relations between $n$ and $H$.
(a), (b), (c), (d) and (e) were given by fixing 
$\kappa\eta = 0.760, 1.15, 1.50, 1.80, 2.01$ respectively.
Values the normal Friedmann equation holds are also indicated at $n = 0$.
As the number of extra-dimensions increase, the expansion rate is suppressed.
The end points of each lines are $(n, H) = (n, 0.01)$.} 
\label{fig:nonSingLocationVn}
\end{figure}

Fig. \ref{fig:nonSingLocationVn} shows the relations
between $n$ and $H$ with $\eta$ fixed at some values.
At $n = 0$ in this figure, the values read from the usual 
four-dimensional Friedmann equation, 
$H^2 = \kappa^2 \rho / 3,~\rho = \lambda\eta^4 / 4$ 
are also indicated. This figure shows that the expansion rate determined 
 by the Friedmann equation is suppressed as 
the  number of extra-dimensions increases and
 $H$ vanished at a specific dimension.
This effect may be considered as the dilution of the cosmological constant. 
However, very small but non-zero expansion rate of $H \sim 10^{-33}\ev$
cannot be reproduced without a fine-tuning for  $\kappa\eta$.

\section{Conclusion}

\label{sec:conclusion}

We have solved the Einstein equations and the scalar field equation 
for a global defect in higher dimensional spacetimes ($\geq 6$). 
The defect has a (3+1) dimensional core in $n\geq 2$ extra dimensions. 
We have extended the analysis by Cho and Vilenkin \cite{CV1}  
to other extra dimensions and found numerical solutions without 
``cigar ansatz''. 
We have examined in detail the relation among the expansion rate $H$ 
and the symmetry-breaking scale $\eta$ and the number of extra dimensions $n$ 
for these solutions. We find that even if the extra dimensions do not have 
a cigar geometry, the expansion rate $H$ grows as $\eta$ increases, which is 
opposite to what is needed for the recently proposed mechanism \cite{DGS} 
for solving the cosmological constant problem.
Finally we want to notice that our nondiverging solutions
require fine-tuning of parameters $(n, \kappa\eta, (\kappa/\lambda^{1/2})H)$.
So the problem of fine-tuning remain exist even if $H$ decreases
as $\eta$ increases.

\begin{acknowledgments}

It is a pleasure to thank Takashi Nakamura for stimulating discussions
throughout the course of this work.
We acknowledge our colleagues of the Kyoto University
for hospitality. 
This work was supported in part by a Grant-in-Aid for Scientific 
Research (No.15740152) from the Japan Society for 
the Promotion of Science,
and also supported by a Grant-in-Aid for the 21st Century COE "Center for Diversity and Universality in Physics''.
 
\end{acknowledgments}

\appendix

\section{Numerics}

\label{sec:numerical}

In this Appendix, we give the details for finding the numerical solutions. 

\subsection{$(\kappa\eta)^2 \leq n-2$ Case}

In this case, our strategy for the calculation is divided into 2 steps.
\begin{enumerate}
\item Firstly numerically integrating Eq. (\ref{eq:EH1}), 
(\ref{eq:EH3}) and (\ref{eq:EOM1}) as an initial value problem from the origin. 
\label{enum:nonSing1}
\item Then solving them as a two-point value problem by the relaxation 
method with the initial solutions obtained by the previous step. 
\label{enum:nonSing2}
\end{enumerate}
In the Step \ref{enum:nonSing1}, we solve the differential equations
by  the 4th-order Runge-Kutta method. 
The calculation starts from the origin with the initial conditions, 
$B(0) = C(0) = 1$, $B'(0) = C'(0) = 0, \phi(0) = 0$ and different $\phi'(0)$. 
The calculation is so sensitive to the initial condition, $\phi'(0)$, 
that bad choice would lead to the divergence of $B,C,\phi$ at finite distance.
So we have fine tuned the sixth initial condition $\phi'(0)$ 
so that the point of divergence goes as far away as possible 
which we call $r_{\text{lim}}$. 
Then we can obtain an approximate solution in the range $[r_{\text{lim}}$, 
$r_{\text{max}}]$ without numerical divergence
by fixing $\phi(r) = \phi(r_{\text{lim}})$ at $r \geq r_{\text{lim}}$ 
artificially. In the Step \ref{enum:nonSing2}, we solve the differential 
equations by the relaxation method with two point boundary conditions at 
$r=0$ and at $r_{\text{max}}$. The sixth condition is now replaced 
with $\phi'(r_{\text{max}}) = 0$. After the iteration converges 
(typically relative error below $5\times 10^{-10}$), we finally obtain a 
solution.

\subsection{$(\kappa\eta)^2 > n-2$ Case}

\begin{figure}
\postscript{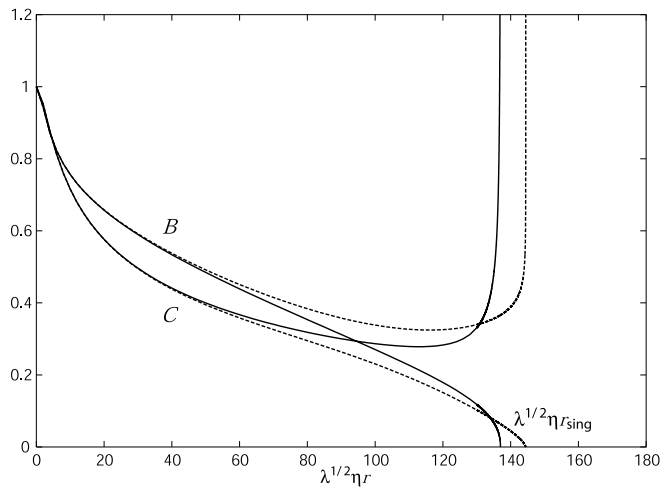}{1.00}
\caption{This graphs shows solutions which have singularity.
The solid line and the broken line is  for the parameter 
$(n, \kappa\eta, (\kappa/\lambda^{1/2})H) = (3, 1.09, 0.00375), 
(3, 1.09, 0.00400)$, respectively.
These solutions have singularity at finite $r$.} 
\label{fig:graph3109n}
\end{figure}

For $(\kappa\eta)^2 > n-2$, arbitrary $H$ including $H = 0$
leads to a divergence in $B$ or $C$ at
finite distance from the origin and the singularity is formed.
We call this point $r_{\text{sing}}$.
At the specific value of $H$, the distance to the singularity becomes as far 
as possible and the divergence vanishes. We call this point $r_{\text{f}}$.

We solve the Einstein equations and the equation of motion of $\phi$ 
numerically in the range of $[0,r_{\text{sing}}]$. 
But the value of $r_{\text{sing}}$ remains unknown to be decided by solving 
accurately. So, our strategy of the calculation 
is divided into 4 steps.
\begin{enumerate}
\item Firstly numerically integrating Eq. (\ref{eq:EH1}), 
(\ref{eq:EH3}) and (\ref{eq:EOM1}) as an initial value problem from the origin. 
\label{enum:Sing1}
\item Then solving them as a two-point value problem by the relaxation 
method with the initial solutions obtained by the previous step. 
\label{enum:Sing2}
\item Extending the solution of $\phi$ to sufficiently large value of $r$. \label{enum:Sing3}
\item Numerically integrating the equations with the fixed $\phi$ obtained
in the Step \ref{enum:Sing3} to find 
the true position of the singularity and the solution. \label{enum:Sing4}
\end{enumerate}
 Step \ref{enum:Sing1} is executed in the same way
as Step 1 of $(\kappa\eta)^2 \leq n-2$ case.
But the range of $r$ is $[0, {\bar r}_{\text{sing}}]$,
where the bar means that this position of the singularity is different 
from the true one. Step \ref{enum:Sing2} is also executed in the same way 
as $(\kappa\eta)^2 \leq n-2$ case. But the boundary condition is imposed at 
$r=0$ and at $r= {\bar r}_{\text{sing}} - \Delta r$. The right point is shifted to 
left by $\Delta r$ so that the singularity is not
included in this range $[0,{\bar r}_{\text{sing}}-\Delta r]$.
By executing the Step \ref{enum:Sing2}, 
solutions of $B, C, \phi$ can be obtained and 
$\phi'({\bar r}_{\text{sing}} - \Delta r)$ is vanishing.
So we can extend $\phi(r)$ to sufficiently large value of $r$,
this procedure is the Step \ref{enum:Sing3}. 
In the Step \ref{enum:Sing4}, we solve the differential equations
by treating the $B(r)$ and $C(r)$ as unknown functions and $\phi(r)$ as 
fixed background obtained in Step \ref{enum:Sing3}.
The algorithm used is the same as Step \ref{enum:Sing1}.
{}From the Step \ref{enum:Sing1} to \ref{enum:Sing4}, we finally 
obtain the numerical solutions and the point of the singularity.

%Finding the eigenvalue.%
Next, we find the "eigenvalue" under the condition that
the position of the singularity becomes as far as possible.
We can find the unique $H$ as the eigenvalue with $(n, \kappa\eta)$ fixed 
like as follows.  If $H$ is smaller than the eigenvalue, the energy of 
the defect core bends and closes the bulk space such that $B(r)\to 0$ as 
$r \to r_{\text{sing}}$, which is shown as solid lines in 
Fig. \ref{fig:graph3109n}. Then $C(r)$ diverge at the same point and 
forming the singularity. As $H$ is increased, the energy density of the defect 
is consumed to inflate the brane and the bulk's curvature is relaxed.
If $H$ is beyond the specific value, then 
that $B(r) \to \infty$ at finite distance $r_{\text{sing}}$, which is 
shown as broken lines in Fig. \ref{fig:graph3109n}.
At a very specific value of $H$ which exists between these two values, 
neither $B(r)$
nor $C(r)$ diverges and the position of the singularity has a local peak 
here. This $H$ is to be called the "eigenvalue". Typical examples are 
given in Fig. \ref{fig:graph3109} and Fig. \ref{fig:combined}.

\section{Cigar type solutions}

\label{sec:cigar}

\begin{figure}
\postscript{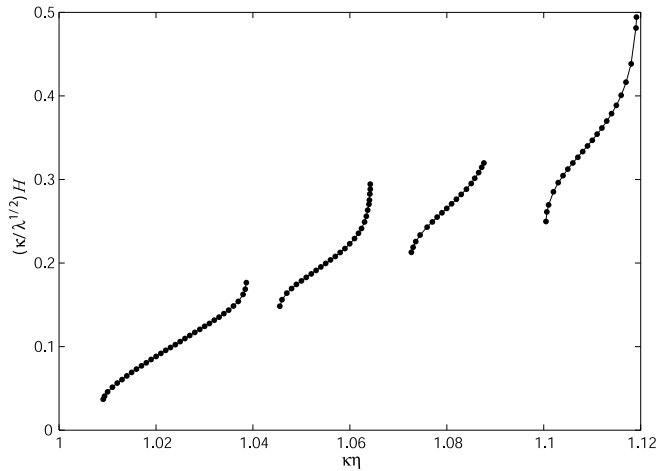}{1.00}
\caption{The relation between $\eta$ and $H$ for cigar type solutions
with $n = 3$.} 
\label{fig:nonSingLocationC3}
\end{figure}
In the previous works \cite{CV1,CV2},
the cigar type solutions in $n = 3$ case are studied.
These solutions have asymptotic forms 
$\sqrt{\lambda}\eta Cr \to \text{constant}$.
The analytic solution for arbitrary $n \geq 3$ is 
\begin{equation}
\kappa^2\phi^2(r_{\text{max}}) = \frac{2(n^2-4) - (n-1)(\kappa\eta)^2}{(n+5)},
\end{equation}
\begin{equation}
B(r_{\text{max}}) = \frac{H}{\sqrt{\lambda}\eta k}
\sin (\sqrt{\lambda} \eta k r_{\text{max}}),
\end{equation}
\begin{equation}
\lambda\eta^2 C^2(r_{\text{max}})r_{\text{max}}^2 = 
\frac{(n-1)(n+5)(\kappa\eta)^2}{2(n+2)[(\kappa\eta)^2-(n-2)]},
\end{equation}
where
\begin{equation}
k = \sqrt{\frac{n+2}{2(n+5)^2}} \frac{(\kappa\eta)^2 - (n-2)}{\kappa\eta}
\end{equation}
and $r_{\text{max}}$ is a sufficiently large value.

It is concluded in \cite{CV1} that eigenvalues
in $(\kappa\eta, (\kappa/\lambda^{1/2})H)$ space can be line fitted.
But we have reexamined this calculation and found more
complex structure shown in Fig. \ref{fig:nonSingLocationC3}.

Our method of the numerical integration is as follows. 
We have set the seven boundary conditions
$B(0) = C(0) = 1,~B'(0) = C'(0) = 0,~\phi(0) = 0,~\phi'(r_{\text{max}}) 
= 0,~(r_{\text{max}}C(r_{\text{max}}))' = 0$,
and considered the equations (\ref{eq:EH1}),(\ref{eq:EH3}), (\ref{eq:EOM1}) 
and $H'(r) = 0$, in which $H$ is treated as the dependent variable of $r$.
The strategy to solve these equations is the same as described 
in Appendix \ref{sec:numerical}. 
The iteration is not converged for some values of $\kappa\eta$ corresponding to
the blanks in Fig. \ref{fig:nonSingLocationC3}.

%%%%%%%%%%%%%%%%%%%%%%%%%%%%%%%%%%%%%%%%%%%%%%%%%%%%%

\end{document}